\documentclass[reprint,
nofootinbib,
aps,
pra,
showkeys
]{revtex4-2}
\pdfoutput=1

\usepackage{amsthm, amsmath, amssymb, amsfonts, dsfont, mathrsfs}
\usepackage{tensor, braket, physics} 
\usepackage{nicefrac}

\usepackage{orcidlink}
\graphicspath{ {./images/} }

\usepackage{algorithm}
\usepackage{algorithmicx}
\usepackage[noend]{algpseudocode}

\usepackage{graphicx, placeins, float}
\usepackage[caption=false]{subfig}
\usepackage[export]{adjustbox}

\usepackage{comment}

\usepackage{hyperref, xcolor}
\usepackage[nameinlink,capitalize]{cleveref}
\hypersetup{
	colorlinks = true,
	linkbordercolor = {white},
	linkcolor={purple},
	citecolor={purple},
	urlcolor={blue}
}

\DeclareMathOperator{\supp}{supp}

\DeclareMathOperator{\QFT}{QFT}
\DeclareMathOperator{\IQFT}{QFT^{\dagger}}
\DeclareMathOperator{\FFT}{FFT}

\newcommand{\dket}[1]{\ket{#1}\rangle}
\newcommand{\Unit}{\mathds{1}}

\newtheorem*{theorem*}{Theorem}

\newtheorem*{corollary*}{Properties}

\makeatletter
\renewcommand{\ALG@name}{Algorithm}
\makeatother
\crefname{algorithm}{Alg.}{Algs.}

\begin{document}
\title{Quantum JPEG}
\author{Simone Roncallo\,\orcidlink{0000-0003-3506-9027}}
	\email[Simone Roncallo: ]{simone.roncallo01@ateneopv.it}
	\affiliation{Dipartimento di Fisica, Università degli Studi di Pavia, Via Agostino Bassi 6, I-27100, Pavia, Italy}
	\affiliation{INFN Sezione di Pavia, Via Agostino Bassi 6, I-27100, Pavia, Italy}
	
\author{Lorenzo Maccone\,\orcidlink{0000-0002-6729-5312}}
	\email[Lorenzo Maccone: ]{lorenzo.maccone@unipv.it}
	\affiliation{Dipartimento di Fisica, Università degli Studi di Pavia, Via Agostino Bassi 6, I-27100, Pavia, Italy}
	\affiliation{INFN Sezione di Pavia, Via Agostino Bassi 6, I-27100, Pavia, Italy}
	
\author{Chiara Macchiavello\,\orcidlink{0000-0002-2955-8759}}
	\email[Chiara Macchiavello: ]{chiara.macchiavello@unipv.it}
	\affiliation{Dipartimento di Fisica, Università degli Studi di Pavia, Via Agostino Bassi 6, I-27100, Pavia, Italy}
	\affiliation{INFN Sezione di Pavia, Via Agostino Bassi 6, I-27100, Pavia, Italy}

\begin{abstract}
	The JPEG algorithm compresses a digital image by filtering its high spatial-frequency components. Similarly, we introduce a quantum algorithm that uses the quantum Fourier transform to discard the high spatial-frequency qubits of an image, downsampling it to a lower resolution. This allows one to capture, compress, and send images even with limited quantum resources for storage and communication. We show under which conditions this protocol is advantageous with respect to its classical counterpart.
\end{abstract}
\maketitle

\section{Introduction}
Digital images represent information in terms of arrays of pixels. Compression and downsampling can reduce the cost for data storage and transmission, while preserving the original visual pattern \citep{book:Solomon}. An example is the joint photographic expert group (JPEG) algorithm \citep{rep:JPEG}, which operates in the spatial-frequency domain. The JPEG algorithm divides the input image into smaller subimages, then taking the discrete cosine transform of each element. The high spatial frequencies are removed, reducing the amount of information stored at the cost of the quality of the output image.

\begin{figure}[b]
	\centering
	\includegraphics[width = 0.45 \textwidth]{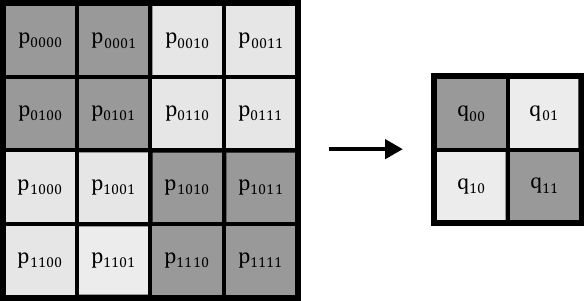}
	\caption{\label{fig:CompressionIntroduction}Downsampling of a $4 \times 4$ image into a $2 \times 2$ image. The input is encoded in the probabilities of a $n_0 = 4$ qubits state $\ket{\Psi}_0$, i.e. $p_{abcd} = |\bra{abcd} \ket{\Psi}_0|^2$ with $a,b,c,d = 0,1$. \cref{alg:Algorithm1} compresses $\ket{\Psi}_0$ into the $n_2 = 2$ qubits state $\rho_2$, yielding the output image from the Born rule $q_{ab} = \Tr(\ket{ab}\!\bra{ab}\rho_2)$.}
\end{figure}

Quantum image processing seeks to encode, manipulate, and retrieve visual information in a quantum-mechanical way \citep{art:Wang, book:Yan,art:Yan,art:Le,art:Venegas,art:Zhang_NEQR, art:Yao}. In this paper, we discuss the downsampling and compression of images encoded as multiqubit quantum states. Using the quantum Fourier transform ($\QFT$) \citep{book:Nielsen,art:Hales}, we provide an algorithm to discard the most significant qubits in the encoding register, thus filtering the high spatial-frequency components of the input image. This algorithm preserves the original visual pattern, downscaling its resolution and reducing the number of encoding resources in the register (see \cref{fig:CompressionIntroduction}). Our implementation differs from \citep{art:Latorre,art:Jiang,art:Zhou,art:Yan_Interpolation} in both the encoding and the compression strategies. In particular, \citep{art:Latorre} leverages matrix product states truncation, while \citep{art:Jiang} and \citep{art:Zhou,art:Yan_Interpolation} propose hybrid or interpolation-based algorithms, respectively. It also differs from \citep{art:XiaotingWang} (appeared at the same
time as the current paper), which focuses on image filtering rather than downsampling.

By taking into account the statistical reconstruction at the output, we show that our algorithm is advantageous over its classical counterpart, as long as the output resolution is sufficiently compressed. As a possible encoder implementation, we consider a multiatom lattice sensor that electromagnetically interacts with a multimode coherent state, encoding an image as the probabilities of a multiqubit quantum register. Our results use the amplitude encoding scheme \citep{book:Petruccione}, but they are completely independent of the hardware model.

\section{Quantum image downsampling and compression\label{sec:II}}
In this section we discuss the quantum downsampling and compression algorithm. We vectorize and encode the image in the probabilities of a state loaded in the $n_0$-register, i.e. a register of $n_0$ qubits. After a $\QFT$, we trace out the qubits that correspond to the high spatial-frequency components of the image. Then, we return to the computational basis with the inverse $\QFT$ ($\IQFT$), discarding the redundant qubits and compressing the image in the $n_2$-register, i.e. a register of $n_2 < n_0$ qubits. We discuss the relation between the downsampling parameter, i.e. the number of discarded qubits $n_2 - n_0$, and the loss in the output resolution.

A digital grayscale image is a grid of elements called \emph{pixels}. Each pixel is associated with a finite and discrete quantity called \emph{gray value}, which represents the brightness reproduced at each point of the grid (for colored images, there are three brightness values, one for each of the red, green and blue channels). Classically, each pixel value is encoded in a string of $c$ bits which determines the total number of $L = 2^c$ \emph{gray levels} of the image, namely the depth. Mathematically, an image is described by a matrix $\mathscr{I}$, with the size $N \times M$ representing its \emph{resolution} and $NM$ the total number of pixels. For simplicity we consider only square images, i.e. with $N=M$.

We represent $\mathscr{I}$ as a $n_0$-qubit quantum state,
\begin{align}
	\ket{\Psi}_0 = \sum_{j=0}^{{2}^{{n}_{0}}-1}  \sqrt{\theta(\mathscr{I}')_j} \ket{j}_{0} \ ,
	\label{eq:InitialVectorizedEncoding}
\end{align}
where $\ket{j}_{0}$ labels the elements of the computational basis on the $n_0$-qubit Hilbert space $\mathcal{H}_0$ and $\theta$ is a row-wise\footnote{The choice of the vectorization path is conventional, as long as it is respected throughout the algorithm.} vectorization map that associates a $N \times N$ matrix to the $N^2$ column vector
\begin{equation}
	\theta(\mathscr{I}) = \left[\mathscr{I}_{00}, \ldots  , \mathscr{I}_{0(N-1)} , \mathscr{I}_{10} , \ldots , \mathscr{I}_{(N-1)( N-1)}\right]^T \ .
\end{equation}
We call $\ket{\Psi}_0$ quantum image. In our notation, states or operations with subscript $0$ are referred to the $n_0$-register. This representation encodes the gray value of $N^2$ pixels in the probabilities of $n_0 = 2\log_2(N)$ qubits, whose normalization follows by rescaling each pixel value to the total brightness of the image, so that $\mathscr{I}' = \mathscr{I} / \sum_{m,n=0}^{N-1}\mathscr{I}_{mn}$. This strategy yields a first lossless exponential compression of the image: from $N^2$ pixels to $n_0$ qubits. Below, we process the quantum image to further reduce the size of the encoding register.

The $\QFT$ operates on the $n_0$-register as
\begin{equation}
	U_{\QFT_{0}} \ket{j}_{0} = \frac{1}{N}\sum_{k=0}^{N^2-1} e^{2\pi ijk/N^2}\ket{k}_{0} \ .
\end{equation}
We use the subscript notation to specify the register on which the $\QFT$ operates. Consider the $n_1$-subregister made only of the first $n_1 = n_0 - \tilde{n}$ qubits of $\mathcal{H}_{0}$ and let $\ket{j}_{1}$ label the elements of its computational basis. Then $\ket{j}_{0} = \ket{l}_{\tilde{n}} \otimes \ket{m}_{1}$, with $\ket{l}_{\tilde{n}}$ defined on the $\tilde{n}$-register. The inverse $\QFT$ operates naturally on the $n_1$-subregister as $(\Unit \otimes U^{\dagger}_{\QFT_{1}}) \ket{j}_{0} = \ket{l}_{\tilde{n}} \otimes U^{\dagger}_{\QFT_1} \ket{m}_{1}$.

Our algorithm downsamples a quantum image $\ket{\Psi}_0$ from the $n_0$-register to the $n_2$-register, with settable $\tilde{n}$, reducing the number of encoding qubits from $n_0$ to
$n_2 = n_0 - 2\tilde{n}$. The output image resolution is $N/2^{\tilde{n}} \times N/2^{\tilde{n}}$. 

We adopt the little-endian ordering, with the least significant qubit placed on the top of the register and labeled by $q=0$, with $0 \leq q \leq m - 1$ and $m \in \{ n_0, n_1, n_2 \}$.\footnote{Consider the 3-qubit state $\ket{abc} = \ket{a} \otimes \ket{b} \otimes \ket{c}$, with $a,b,c = 0,1$. The little-endian convention interprets $a$ as the most significant qubit in the register. In our notation, it is labeled by $q=3$.} We denote $H$ the single-qubit Hadamard gate.
\begin{algorithm}[H]
\caption{\label{alg:Algorithm1}Quantum image downsampling}\vspace{2pt}
\textbf{Input} Quantum image $\ket{\Psi}_0$ \\[1pt]
\textbf{Parameter} Integer $\tilde{n} < n_0/2$ 
\begin{algorithmic}[1]
\State apply $H^{\otimes n_0}$ \Comment{$n_0$-register}
\State apply $U_{\QFT_0}$ \vspace{2pt}
\For{$q$ in $n_0$-register}
	\If{$n_0 - \tilde{n} \leq q \leq n_0 - 1$}
		\State{discard the $q$th qubit} \Comment{Rule 1 ($n_0 \to n_1)$}
	\EndIf
\EndFor
\State apply $U^{\dagger}_{\QFT_1}$ \Comment{$n_1$-register}
\For{$q$ in $n_1$-register}
	\If{$n_0 / 2 - \tilde{n} \leq q \leq n_0 / 2 - 1$}
		\State{discard the $q$th qubit} \Comment{Rule 2 ($n_1 \to n_2)$}
	\EndIf
\EndFor
\State apply $H^{\otimes n_2}$ \Comment{$n_2$-register}
\end{algorithmic}
\textbf{Result} Compressed image $\rho_2$ 
\end{algorithm}
The algorithm operates as follows. After applying $H^{\otimes n_0}$, a first $\QFT$ is taken on the register in which $\ket{\Psi}_0$ is initially loaded. Discard the first $\tilde{n}$ most significant qubits from the bottom of the register, which means averaging out the low-probability frequency components of the original image (Rule 1). Take the $\IQFT$. This reduces the image resolution of a factor $2^{\tilde{n}}$ along only one of its axes. Then, discard the last $\tilde{n}$ positions of the first half of the initial register, which contain redundant information after the $\IQFT$, then take $H^{\otimes n_2}$. As we discuss below, this yields a $N/2^{\tilde{n}} \times N/2^{\tilde{n}}$ image, conformally downscaled along both its axes, and described by the state $\rho_2$. There may be different implementations of this protocol, which eventually uses the superposition principle to achieve a Fourier compression that operates on both the image axes, simultaneously. However, we expect such generalizations to produce similar results, eventually optimizing the complexity of the whole algorithm or maximizing the signal-to-noise ratio of the final reconstruction. 

\begin{figure}[t]
	\centering
	\includegraphics[width = 0.48 \textwidth]{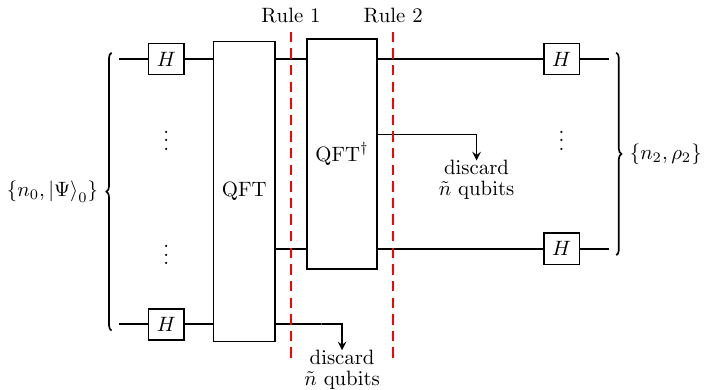}
	\caption{\label{fig:Algorithm1}Downsampling of a quantum image into a register of fewer qubits. The image is initially encoded in the probabilities of $\ket{\Psi}_0$. The most significant qubit is represented at the bottom of the $n_0$-register. The $\QFT$ is taken. (Rule 1) Chosen an integer $\tilde{n} < n_0 / 2$, take an $\IQFT$ on the first $n_0 - \tilde{n}$ qubits and discard the remaining part of the register. (Rule 2) Discard the same number of qubits from those register positions that correspond to the last $\tilde{n}$ qubits of the first half of the $n_0$-register. The algorithm compresses the image $\ket{\Psi}_0$ in the state $\rho_2$, reducing the number of pixels from $N^2$ to $N^2/4^{\tilde{n}}$. The Hadamard gates reduce the statistical fluctuations at the output, while preserving the original image contrast (see \cref{app:A} for a discussion). The trade-off between the output resolution and the amount of resources saved (the number of discarded qubits) is controlled by the choice of $\tilde{n}$.}
\end{figure}

The algorithm works independently of the presence of $H^{\otimes n_0}$ (Step 1) and $H^{\otimes n_2}$ (Step 10). As we show in \cref{app:A}, the Hadamard gates improve the quality at the output, reducing the statistical fluctuations at each pixel while preserving the original image contrast. We report the circuit implementation of \cref{alg:Algorithm1} in \cref{fig:Algorithm1}.

As an example of Rule 1 and 2, consider a $512 \times 512$ image encoded in a $18$-qubit register, with qubits labeled by the list $[0,1,\ldots,17]$. Suppose that $\tilde{n} = 4$. The discarding sequence reads
\begin{multline}
	[0,1,\ldots,17] \xrightarrow[\text{Discard } {[}14,15,16,17{]}]{\text{Rule 1}} [0,1,\ldots,13] \\
	 \xrightarrow[\text{Discard } {[}5,6,7,8{]}]{\text{Rule 2}} [0, 1, 2, 3, 4, 9, 10, 11, 12, 13] \ .
\end{multline}
These rules, combined with the $\QFT$ and $\IQFT$, yield a downsampled $32 \times 32$ image encoded in a $10$-qubit register, with qubits $[0, 1, 2, 3, 4, 9, 10, 11, 12, 13]$.

The reconstruction of the output image requires the complete knowledge of the probabilities
\begin{equation}
	p_j = \Tr(\ket{j}_2 \! \bra{j} \rho_2) \ ,
	\label{eq:OutProb}
\end{equation}  
with $j \in \{0,1,\ldots,2^{n_2}-1\}$. The image is obtained by means of a devectorization operation $\theta^{-1}$, which rearranges $p_j$ as the entries of a $d \times d$ matrix, with $d^2 = 2^{{n}_{2}}$. Characterizing the sample size for a good-quality output requires further statistical considerations.

\begin{figure}
	\centering
	\subfloat[]{\includegraphics[width = 0.22 \textwidth]{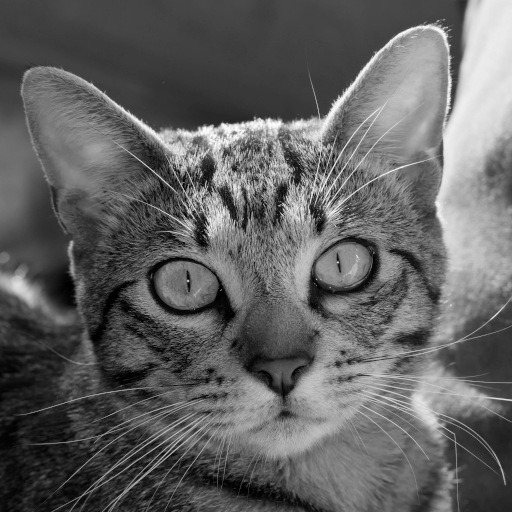}}\hspace{0.01 \textwidth}%
	\subfloat[]{\includegraphics[width = 0.22 \textwidth]{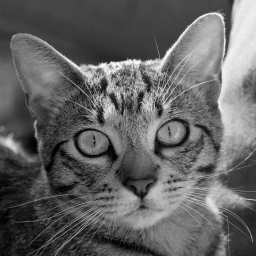}}
	
	\subfloat[]{\includegraphics[width = 0.22 \textwidth]{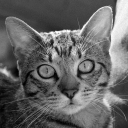}\label{fig:FluctuationSample}}\hspace{0.01 \textwidth}%
	\subfloat[]{\includegraphics[width = 0.22 \textwidth]{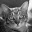}}
	
	\subfloat[]{\includegraphics[width = 0.5 \textwidth]{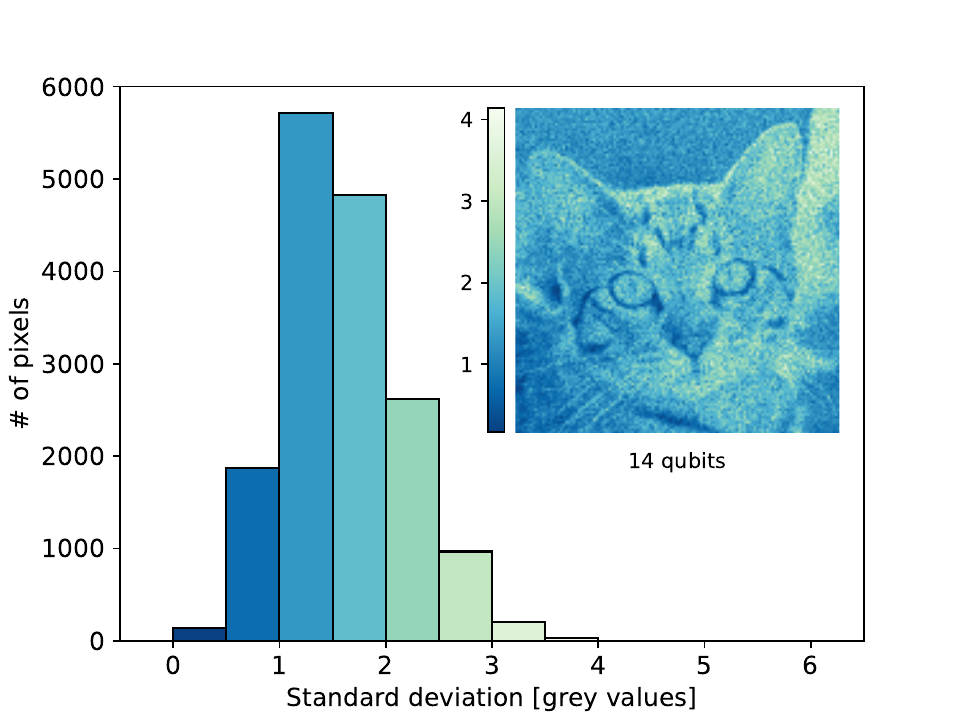}\label{fig:Fluctuations}}
	\caption{\label{fig:Simulation1}Simulated results for a $8$-bit digital image with original resolution of $512 \times 512$ pixels. (a) The image is initially vectorized and prepared as a $18$-qubit quantum state, then \cref{alg:Algorithm1} is taken. Figures (b-d) show the reconstructed output with respect to the number of $n_2$ qubits in the output register. The Hadamard gates improve the reconstruction, reducing the statistical fluctuations while correcting the image contrast (see \cref{app:A} for a discussion). Simulation performed with Qiskit Aer and $2^{n_2} \times 256^{\nicefrac{3}{2}}$ shots. For comparison, all the images are rescaled to the same size. Aliasing starts to appear on cat's whiskers. (b) $n_2 = 16$, $256 \times 256$ pixels. (c) $n_2 = 14$, $128 \times 128$ pixels. (d)  $n_2 = 10$, $32 \times 32$ pixels. (e) Statistical fluctuations at the output. The reconstruction of (c) is repeated for $20$ times. The histogram plots the number of pixels with respect to their standard deviation. The inset shows the standard deviation at each pixel location.}
\end{figure}

Since the gray values are probabilistically encoded in the output state, a single computational basis measurement on $\rho_2$ would produce one white pixel in the image plane, with position specified by the output bitstring. The complete reconstruction of the image requires the full knowledge of the probability distribution of \cref{eq:OutProb}. Let $S$ be the number of shots, obtained by repeating the algorithm and the measurement for $S$ times, and $L$ be the number of gray levels desired at the output, e.g. $L = 256$ for an $8$-bit image. Let $f_j$ be the frequency of each outcome. We identify the color \emph{white} with the pixel with the highest probability $f_{w} = \max_j f_{j}$, with  $f_{w} \in [0,1]$ and $f_j \in [0, f_{w}] \ \forall j \in \{0, 1, \ldots, d^2 -1\}$. Each gray value is reconstructed as 
\begin{equation}
	g_j = \frac{f_jL}{f_{w}} \ ,
\end{equation}
with $g_j = 0, 1, \ldots L-1$ and $g_{w} = L$. A measurement on $\rho_2$ corresponds to two possible outcomes: a shot is assigned to the $j$th bin, i.e. to the $j$th pixel, with probability $p_j$, or it is not with probability $1-p_j$. Under the normal approximation and with the $95\%$ confidence level \citep{book:Rotondi}, the probabilities can be estimated as $p_j = f_j \pm 2\sqrt{f_j(1-f_j)/S}$. In terms of gray values, this reads
\begin{equation}
	g_j = \frac{f_jL}{f_{w}} \pm 2 \sqrt{\frac{f_j(1-f_j)L^2}{f_{w}^2 S}} \ .
\end{equation}
By requiring that $g_j$ fluctuates of at most one gray level, we get $S \geq 4f_j(1-f_j)L^2/f_{w}^2$. Any realistic image with at least two non-black pixels has $f_{w} < 0.5$, yielding $S \geq 4L^2/f_{w}$. For an all-purpose estimation, consider a $d \times d$ completely white image as the worst-case scenario. In this case $f_w = 1/d^2$ and
\begin{equation}
	S \simeq 4L^2d^2  \ ,
	\label{eq:SampleSize}
\end{equation}
where each pixel uniformly collects an average of $4L^2$ shots. 

\cref{eq:SampleSize} provides a conservative estimation for every image. The optimal sample size depends on the proportion between dark and bright pixels, which determines the value of $f_w$.\footnote{An example is given by highly contrasted and mostly-black images, in which small samples of data will quickly localize in the most significant sectors of the image.} For a specific image, the reconstruction can be actively optimized by measuring the fluctuations each time a new shot is collected, then stopping the experimental repetitions as soon as the desired standard deviation is reached by the chosen amount of pixels.

In terms of number of gates, \cref{alg:Algorithm1} provides an exponential speedup over its classical counterpart. However, the output reconstruction requires multiple runs of it, increasing the overall cost of the protocol. We show that an advantage remains, which increases when registers of different sizes are all downsampled to the same resolution. Consider a $N \times N$ ($n_0$-qubit) image and the $N /2^{\tilde{n}} \times N /2^{\tilde{n}}$ ($n_2$-qubit) output of \cref{alg:Algorithm1}, with $L = 2^c$ gray levels and downsampling parameter $\tilde{n} = (n_0 - n_2)/2$. For $n_1 \ll n_0$, the complexity of $\QFT_0$ dominates that of $\IQFT_1$, while for $n_1 \simeq n_0$ both contribute to the same order. For this reason, their composition can be upper bounded at $\mathcal{O}(2 n_0^2)$ gates.\footnote{This cost can be upper bounded at $\mathcal{O}(2 n_0 \log_2(n_0))$, by instead approximating the $\QFT$ \citep{art:Hales}.} A conservative reconstruction requires $4L^2 2^{n_2}$ shots, for a total cost of $Q(n_0, n_2) = \mathcal{O}(8 L^2 n_0^2 2^{n_2})$ operations.\footnote{See \cref{eq:SampleSize}. Notice that the total cost depends on a trade-off between the input and the output resolutions, i.e. on $n_0$ and $n_2$, which determines the advantage.} A single fast Fourier transform ($\FFT$) requires $\mathcal{O}\left(2 N^2 \log_2N \right)$ operations \citep{art:Brigham}, for a total classical cost $C(n_0) = \mathcal{O}\left(2 n_0 2^{n_0} \right)$. Then, the relative cost in terms of operations and statistics is $Q(n_0,n_2)/C(n_0) = \mathcal{O}(n_02^{2+2c+n_2-n_0})$. An advantage holds whenever $Q(n_0,n_2) < C(n_0)$, namely
\begin{equation}
	2(1+c) + \log_2(n_0) < 2\tilde{n} < n_0 \ .
	\label{eq:AdvantageInequality}
\end{equation} 
For example, downsampling a $128 \times 128$ ($14$-qubit) black-and-white image to a $8 \times 8$ ($6$-qubit) resolution requires $\sim 12 \%$ fewer operations using \cref{alg:Algorithm1} than the $\FFT$. 

In \cref{fig:Simulation1}, we simulate the downsampling and reconstruction of a $8$-bit $512 \times 512$ image, prepared in a $18$-qubit register. The output is shown for different values of $n_2$. In \cref{fig:Fluctuations}, we plot the statistical fluctuations for a reconstruction performed with $L^{3/2}d^2$ shots. We show that this order of magnitude provides the adequate statistics for an average reconstruction.

The vectorized representation of \cref{eq:InitialVectorizedEncoding} allows the processing of high-resolution images even for small NISQ registers \citep{art:Preskill_NISQ}. For example, let $b$ be the maximum number of qubits supported by an hypothetical hardware device, and consider a $N \times N$ image exceeding the encoding capability of the hardware with $N^2 > 2^{b}$. The compression can still be achieved by first splitting the image in $2^{b/2} \times 2^{b/2}$ subimages, then processing it as a sequence of vectorized quantum states $\left\{\ket{\Psi^{(m)}}_0\right\}_m$, each given in input to  \cref{alg:Algorithm1}, with $m = 0,1, \ldots, N^2/2^{b}$. A similar pre-encoding is also performed by the classical JPEG algorithm, which splits the original image in $8 \times 8$ blocks, before mapping them to the spatial frequencies domain \citep{rep:JPEG}.

Finally, we outline an alternative algorithm that compresses the image without downscaling its resolution. Consider $U_{\QFT_0}\ket{\Psi}_0$. Instead of discarding the last $\tilde{n}$ qubits of the $n_0$-register, we apply Rule 1 to reinitialize them to $\ket{0 \ldots 0}_{\tilde{n}}$.\footnote{This operation can be unitarily implemented by adding as many ancillas and SWAP operations as the number of qubits to be reinitialized.} After taking $\IQFT_0$ and skipping Rule 2, we obtain $\rho_0$, without modifying the size of the original register. This implementation closely relates to the classical JPEG compression, in particular when combined with the previous subimages encoding. As opposed to \cref{alg:Algorithm1}, this algorithm preserves the resolution of the input image. For this same reason, however, the full output reconstruction shows no computational advantage with respect to its classical counterpart.

\section{Hardware encoder: The quantum camera\label{sec:III}}
In this section, we show how to encode an image into the $n_0$-register using a multiatom lattice sensor analogous to a classical charge-coupled device (CCD) \citep{art:Boyle}. This provides a hardware alternative to the Grover-Rudolph preparation scheme \citep{art:Grover-Rudolph}. The implementation below precedes the application of \cref{alg:Algorithm1}, whose working principle remains independent of the specific hardware encoder model or the preparation strategy. The output of the sensor is related to the $n_0$-register of \cref{alg:Algorithm1}, in which the fingerprint of the image is loaded as the multiqubit quantum state $\ket{\Psi}_0$.

Consider a classical image encoded in a multimode coherent state $\ket{\alpha_{\psi}} = D(\alpha_\psi)\ket{\varnothing}$, where $\ket{\varnothing}$ denotes the vacuum state and
\begin{equation}
	D(\alpha_\psi) = \exp(\alpha A^\dagger_\psi - \alpha^* A_\psi)
\end{equation}
the multimode displacement operator. We work in the far-field approximation, with $k$ and $r = (x,y)$ respectively labeling the momentum and the transversal position on the image plane. The creation operator
\begin{equation}
	A_\psi^\dagger = \int d k \widetilde{\psi}(k) a^\dagger(k)
\end{equation}
encodes the Fourier transform $\widetilde{\psi}(k) = \int dr \psi(r) e^{-ik \cdot r}$ of the spectral amplitudes of $\ket{\alpha_{\psi}}$. In our notation, $\psi$ describes the classical analog image encoded in the coherent state, while $\ket{\Psi}_0$ denotes the same image but discretized, vectorized and probabilistically encoded in the $n_0$-register of \cref{sec:II}. In this section, we show how to connect these two representations.

We consider a lattice made of $N^2$ identical atoms, with evenly-spaced energy levels. Let $(m,n)$ label their position on the $N \times N$  lattice. Each atom can be modeled as a two-level system, namely a qubit with computational basis $\left\{\ket{0}_{mn}, \ket{1}_{mn}\right\}$. We initialize the lattice in the vacuum state $\ket{b} = \bigotimes_{m',n'}\ket{0}_{m'n'}$, which represents a completely black image with all the qubits in the ground state.

We model the interaction between the lattice and the electromagnetic field using a multimode and multiatom Jaynes-Cummings Hamiltonian, in the resonant rotating wave approximation \citep{art:Emary_MultiJC, art:Wickenbrock_MultiJC, book:Gerry}
\begin{equation}
	H_{I} = \int dk \sum_{m,n=0}^{N-1} \gamma^*_{mn}(k) a(k)\sigma_{mn}^{+} + \text{ H.c.} \ ,
	\label{eq:JaynesCummingsHamiltonian}
\end{equation}
where
\begin{equation}
	\sigma_{mn}^{+} = \ket{1}_{mn}\!\bra{0} \ , \quad
	\sigma_{mn}^{-} = \ket{0}_{mn}\!\bra{1} \ ,
\end{equation}
and $\gamma_{mn}(k)$ is the Fourier transform of the spatial coupling function $g_{mn}(r)$ between the coherent photons and a qubit located at $(m,n)$, i.e. $\gamma_{mn}(k) = \int dr g_{mn}(r) e^{-i k \cdot r}$. We assume that $g_{mn}$ is compactly supported $\forall m,n$ and that $\supp(g_{mn}) \cap \supp(g_{m'n'})$ is negligible. 

\begin{figure}[t]
	\centering
	\subfloat[]{\includegraphics[width = 0.235 \textwidth , valign = b]{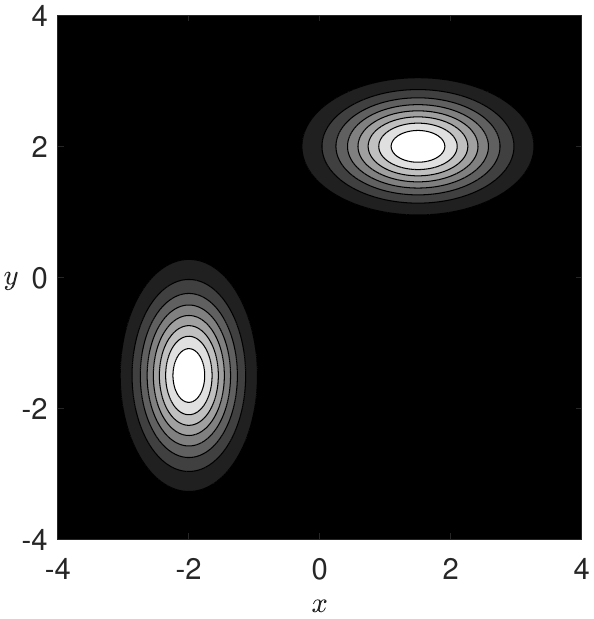}}\hfill%
	\subfloat[]{\includegraphics[width = 0.24 \textwidth , valign = b]{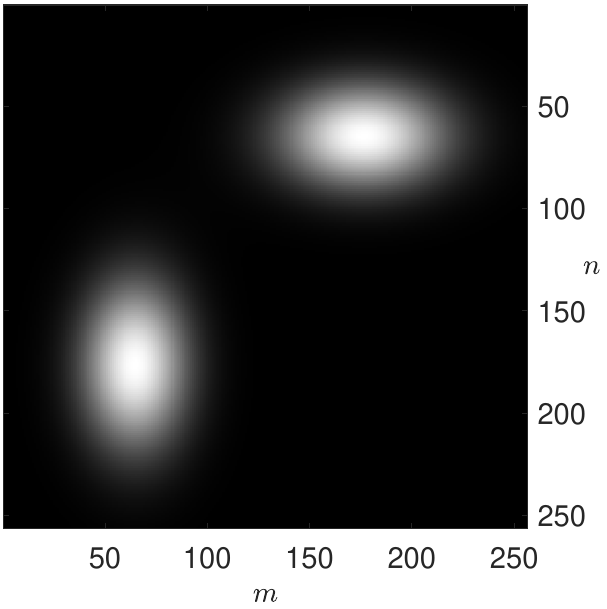}}
	\caption{\label{fig:HardwareEncoding}Comparison between the spectral amplitude $\psi$ and the multiqubit quantum state $\ket{\Omega}$, encoded using the multimode and multiatom Jaynes-Cummings Hamiltonian. (a) Contour plot of $|\psi(x,y)|^2$, obtained from the multimode coherent spatial spectrum $\psi(x,y) = \sum_{i=0}^{1} K \exp(-\alpha_i(x-x_i)^2 - \beta_i(y-y_i)^2)$, with parameters $x_0 = - y_1 = 1.5$, $y_0 = -x_1 = 2$, $\alpha_0 = \beta_1 = 0.35$, $\alpha_1 = \beta_0 = 1$ and normalization constant $K$. (b) Probabilities $|v_{mn}|^2$ of the multiqubit state $\ket{\Omega}$, plotted as a grayscale image, with $m,n = 0,1,\ldots,255$. The sensor is an evenly-spaced lattice of identical two-level systems, with spatial coupling functions $g_{mn}(x,y) = \exp( -5(x-x_m)^2 -5(y-y_n)^2 )$ and positions $(x_m, y_n) \in [-4,4]\times [-4,4]$.}
\end{figure}

The interaction Hamiltonian $H_I$ encodes the input coherent state as a quantum image in the user's sensor. See \cref{app:B} for a complete derivation and a discussion about time evolution. Consider a shutter that initially prevents the photons from interacting with the qubits, so that the initial state reads $\dket{I} = \ket{\alpha_\psi} \otimes \ket{b}$. When the shutter opens, a photon may release energy in the sensor, exciting one of the qubits as $\ket{0}_{mn} \to \ket{1}_{mn}$. Globally, a single interaction may give no excitation, a single excitation or multiple qubits excitations, with contributions to the final state of following the form:
\begin{multline}
	c^0_0 \ket{b} + c^{1}_0\ket{0 \ldots 001} + c^{1}_{1}\ket{0 \ldots 010} + \ldots \\ + c^{1}_{N^2-1} \ket{10 \ldots 00} + c^{2}_{0}\ket{0 \ldots 011} + \ldots 
\end{multline}
The vacuum term can be neglected in post-selection, by discarding any completely black image at the output. Then, the single-qubit contributions become dominant and the final state reads $\dket{F} \simeq \ket{\alpha_\psi} \otimes \ket{\Omega}$, where
\begin{align}
	 \ket{\Omega} = C \sum_{m,n = 0}^{N-1} v_{mn} \sigma^+_{mn} \ket{b} \ , \label{eq:OneHotEncoding} \\
	v_{mn} = \int_{P_{mn}} dr \psi(r) g^*_{mn}(r) \ ,
\label{eq:EncodingCoefficients}
\end{align}
with $P_{mn} = \supp{(g_{mn})}$, and $C$ an overall normalization constant. In these equations, the interaction term $g_{mn}(r)$ acts as a spatial transfer function that discretizes the image in the region $P_{mn}$, encoding the light intensity in the probability of the corresponding qubit to occupy the higher energy level state. In \cref{fig:HardwareEncoding}, we plot the probabilities encoded from a superposition of two Gaussian packets. We compare the numerical results with the probability density of the input state, showing that they correctly reproduce the spectral amplitudes of $\ket{\alpha_\psi}$.

The state $\ket{\Omega}$ represents the image using a sub-optimal one-hot encoding, i.e. with only $N^2$-qubit contributions of the form $(\bigotimes_{m'\neq m,n'\neq n}\ket{0}_{m'n'})\otimes \ket{1}_{mn} \forall m,n \in \{ 0,1,\ldots, N-1 \}$. We optimize the output of the hardware encoder by appending a one-hot to binary converter \citep{art:Chen_Converter,art:Bartschi_Converter}, which uses a combination of CNOTs to re-encode the image in $n_0 = 2\log_2(N)$ qubits,\footnote{This pre-compression prepares the input for \cref{alg:Algorithm1}: It is a \emph{lossless} reorganization of the encoding coefficient $v_{mn}$, inside a smaller but fully exploited quantum register.} yielding
\begin{equation}
	\ket{\Omega} \to \sum_{j=0}^{N^2-1} w_j \ket{j}_0 = \ket{\Psi}_0 \ ,
\end{equation}
which is the state $\ket{\Psi}_0$ of \cref{sec:II}, with $w = \theta(v)$. This conversion requires $\mathcal{O}(N^2)$ operations, which is the same readout cost of the CCD bidirectional shift registers.

After this step, the user applies \cref{alg:Algorithm1} to  discard $2\tilde{n}$ qubits from the encoding register. This compresses $\ket{\Psi}_0$ in $\rho_2$, reducing the requirements for its storage, e.g. the number of qubits of a quantum random access memory \citep{art:Giovannetti_qRAM}, as well as the dimension of the channel needed for communicating the image to another user \citep{art:Gyongyosi}.

\section{Conclusions}
In this paper, we introduced a novel quantum algorithm for image downsampling and compression. Our protocol leverages the $\QFT$ to downscale the original visual pattern, which is preserved while reducing the number of encoding qubits. This opens new perspectives for quantum imaging and quantum image processing. In terms of number of gates only, our scheme provides an exponential speedup over its classical $\FFT$-based counterpart and over interpolation-based quantum algorithms. Although its cost rises when reconstructing full images, we investigated the statistics at the output, showing that the advantage increases consistently with the size of the input register. Whenever a full reconstruction is not needed, e.g. when using our algorithm as a quantum pre-processing operation rather than a standalone module, the theoretical advantage is completely recovered.

As a possible hardware implementation, we designed a multiatom lattice sensor for coherent light. Modeled by the Jaynes-Cummings Hamiltonian, this encoder maps the light intensity at each lattice location to the probability of the corresponding qubit of being excited by a single-photon interaction. In this framework, two users can capture and share images even with limited communication resources, using our downsampling algorithm to forward the image state through lower-dimensional channels.

\section*{Acknowledgments}
This work received support from EU H2020 QuantERA ERA-NET Cofund in Quantum Technologies, Quantum Information and Communication with High-dimensional Encoding (QuICHE) under Grant Agreement 731473 and 101017733, from the U.S. Department of Energy, Office of Science, National Quantum Information Science Research Centers, Superconducting Quantum Materials and Systems Center (SQMS) under Contract No. DE-AC02-07CH11359. L.M. acknowledges support from the PNRR MUR Project PE0000023-NQSTI. C.M. acknowledges support from the National Research Centre for HPC, Big Data and Quantum Computing, PNRR MUR Project CN0000013-ICSC. No animals were harmed in the making of this work. S.R. would like to thank his feline friend for graciously and patiently posing for the picture used in this research.


\section*{Data availability}
The underlying code that generated the data for this study is openly available in GitHub \citep{rep:Quantum_JPEG}.
\appendix

\section{FURTHER CONSIDERATIONS ON THE ALGORITHM\label{app:A}}
In this section, we provide an explicit example of our downsampling algorithm for a simple $4 \times 4$ pattern. We first describe the algorithm without using the Hadamard gates. Then, we show that they reduce the contrast shift and the statistical fluctuations at the output.
\begin{figure*}
	\centering
	\subfloat[]{\includegraphics[width = 0.5 \textwidth]{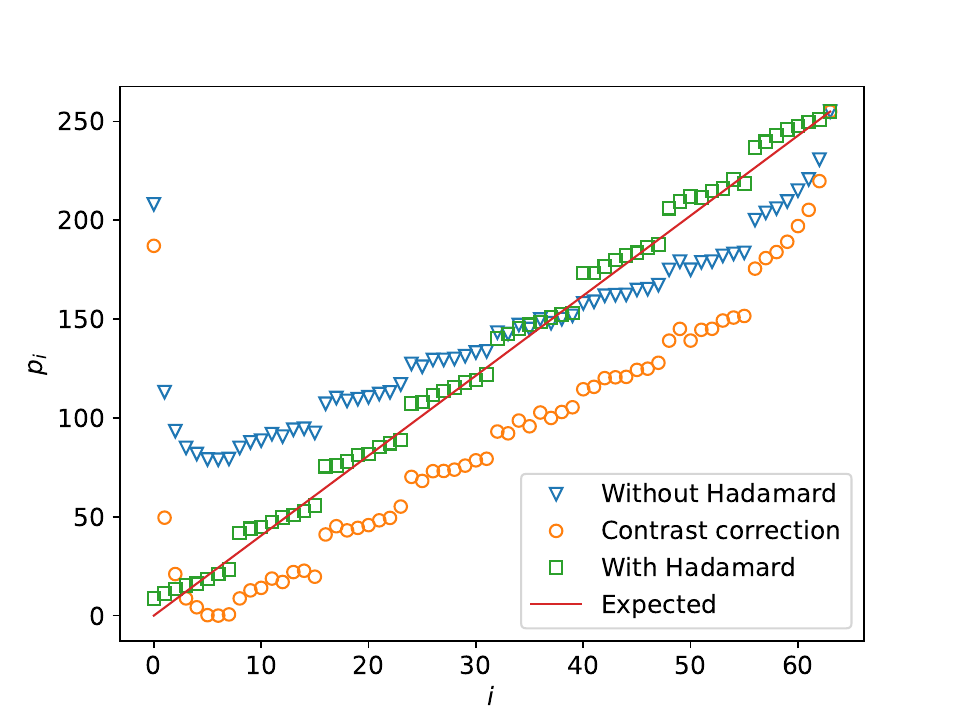}}%
	\subfloat[]{\includegraphics[width = 0.5 \textwidth]{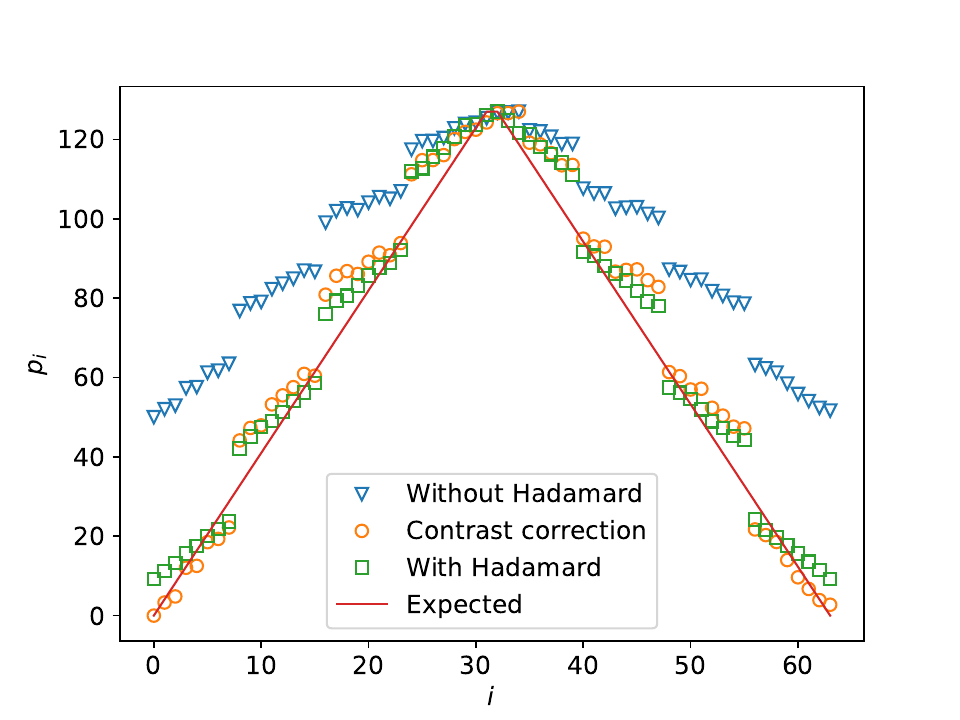}}
	\caption{\label{fig:ComparisonHadamardExample}Downsampling of two arrays prepared as $8$-qubit quantum states, with $\tilde{n} = 1$. The simulation is performed with Qiskit Aer and $10^{6}$ shots. The output of \cref{alg:Algorithm1} is plotted with and without the Hadamard gates, and by post-correcting the contrast in the latter case. The plot shows the probabilities $p_i$ of each multiqubit output state $\ket{i}_{2}$. (a) The input array is $[0,1,2,\ldots,255]$. The continuous line shows the original pattern directly represented in the $6$-qubit register, i.e. the array $[0,1,2,\ldots,63]$. (b) The input is $[0,1,2,\ldots,127,127,126,125,\ldots,0]$. The continuous line shows the same pattern in the $6$-qubit register, i.e. $[0,1,2,\ldots,63,63,62,61,\ldots,0]$. The output best reproduces the original input when using the Hadamard, which preserves the contrast and shows no artifact. The discontinuities at each $8$ multiqubit configuration can be removed by turning off (Rule 2), which remains, however, essential to preserve the aspect ratio when compressing two-dimensional images instead of one-dimensional arrays.}
\end{figure*} 

Consider a lower triangular $4 \times 4$ image
\begin{equation}
	\mathscr{I} = \begin{bmatrix}
		1 & 0 & 0 & 0 \\
		1 & 1 & 0 & 0 \\
		1 & 1 & 1 & 0 \\
		1 & 1 & 1 & 1
	\end{bmatrix} \ .
	\label{eq:ExampleTriangular}
\end{equation}
This pattern can be downscaled into a $2 \times 2$ matrix as
\begin{equation}
	\mathscr{I} = \begin{bmatrix}
		1 & 0 \\
		1 & 1 \\
	\end{bmatrix} \ .
	\label{eq:IdealTriangular}
\end{equation}
We vectorize and represent \cref{eq:ExampleTriangular} as $4$-qubit state
\begin{equation}
	\ket{\Psi}_0 \leftarrow \frac{1}{\sqrt{8}}
	\left[ 1 \ 0 \ 0 \ 0 \ 1 \ 1 \ 0 \ 0 \ 1 \ 1 \ 1 \ 0 \ 1 \ 1 \ 1 \ 1 \right]^T \ .
\end{equation}
With $n_2 = 2$, $\tilde{n} = 1$ and before discarding any qubit, \cref{alg:Algorithm1} reads
\begin{equation}
	 \ket{\Phi}_0 = (\Unit \otimes U^{\dagger}_{\QFT_1})\ U_{\QFT_0}\ket{\Psi}_0 \ .
	 \label{eq:OutputTriangularState}
\end{equation}
We denote the amplitude components of the output state as $\ket{abcd}$, with $a$ ($d$) labeling the bottom (top) qubit on the register of \cref{fig:Algorithm1} and $a,b,c,d = 0,1$. Let $p_{abcd}$ be the probabilities of $\ket{abcd}$. We apply Rule 1 by tracing out the high spatial-frequency qubit $a$ from $\ket{\Phi}_0$, i.e. by summing $p_{0bcd}$ with $p_{1bcd}$ $\forall b,c,d$. Then, we apply Rule 2 by doing the same for the redundant qubit $c$, i.e. by summing $p_{b0d}$ with $p_{b1d}$ $\forall b,d$. The probabilities in the $n_2$-register read
\begin{equation}
	p_{bd} = \sum_{a,c=0}^{1}p_{abcd} \ ,
\end{equation}
which correspond to the $2 \times 2$ image
\begin{equation}
	\mathscr{I}_c = \begin{bmatrix}
		p_{00} & p_{01} \\
		p_{10} & p_{11}
	\end{bmatrix} \ .
\end{equation}
The above procedure is implicitly summarized in \cref{eq:OutProb}, in which the compressed state $\rho_2$ is obtained by applying (Rule 1) and (Rule 2) to $\ket{\Phi}_0$. For \cref{eq:ExampleTriangular}, we find
\begin{equation}
	\mathscr{I}_c \simeq \begin{bmatrix}
		0.35 & 0.06 \\
		0.32 & 0.27
	\end{bmatrix} \ ,
	\label{eq:CompressedTriangular}
\end{equation}
which reproduces the same input pattern, but downsampled and represented with fewer qubits than $\ket{\Psi}_0$.

The results improve when $\ket{\Psi}_0$ is encoded in the $X$ basis. Let $H$ be the single-qubit Hadamard operator. As shown in \cref{fig:Algorithm1}, we introduce $H^{\otimes n_0}$ and $H^{\otimes n_2}$ respectively before $\QFT_0$ and after (Rule 2). Hence, \cref{alg:Algorithm1} reads
\begin{multline}
	\ket{\Psi}_0 \to H^{\otimes n_0} \to U_{\QFT_0} \to \text{Rule 1} \\ \to U^{\dagger}_{\QFT_1} \to \text{Rule 2} \to H^{\otimes n_2} \to \rho_2 \ ,
\end{multline}
yielding
\begin{equation}
	\mathscr{I}_c \simeq \begin{bmatrix}
		0.30 & 0.00 \\
		0.41 & 0.29
	\end{bmatrix} \ ,
\end{equation}
which better reproduces the ideal $2 \times 2$ pattern of \cref{eq:CompressedTriangular}.
\begin{figure*}
	\centering
	\subfloat[]{\includegraphics[width = 0.5 \textwidth]{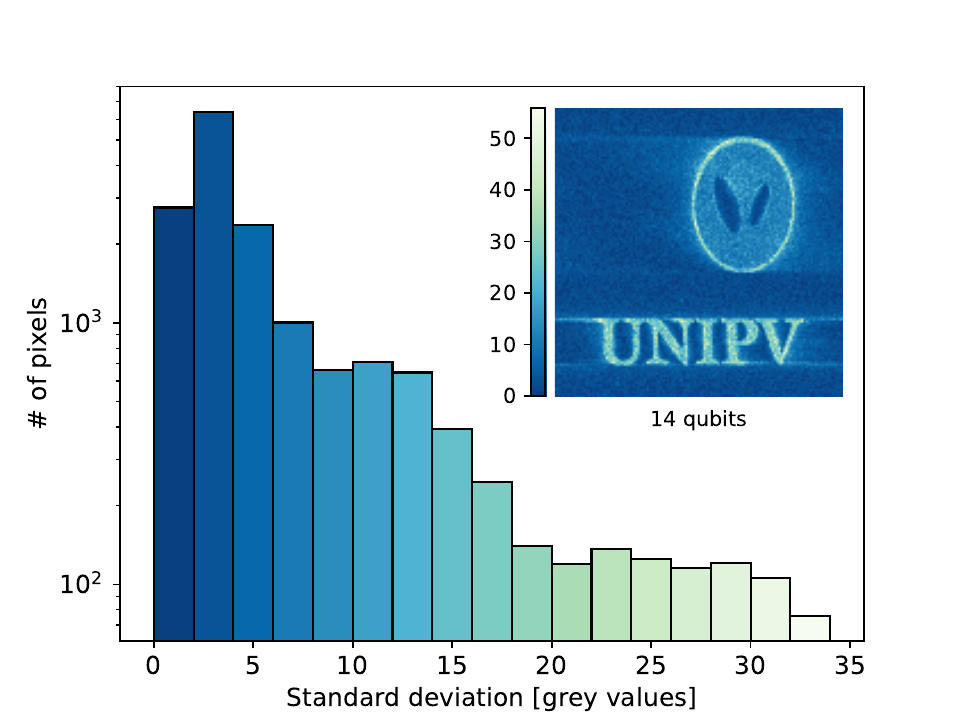}}%
	\subfloat[]{\includegraphics[width = 0.5 \textwidth]{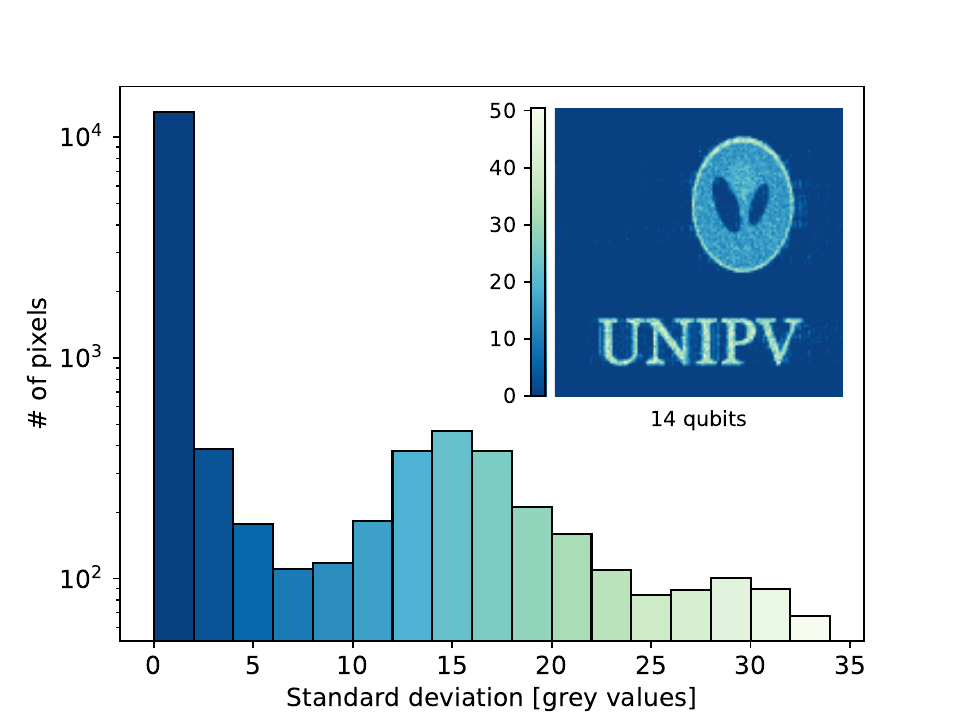}}
	\caption{\label{fig:HadamardComparison}Statistical reconstruction at the algorithm output, without (a) and with (b) the Hadamard gates. The input image represents a Shepp-Logan phantom, commonly used in medical tomography \citep{art:Shepp}, which is prepared as a $18$-qubit quantum state and downsampled for $\tilde{n} = 2$. The simulation is performed with Qiskit Aer and $2^{14}$ shots. The choice of a sub-optimal sample size purposely showcases the effect of the statistical fluctuations. The histograms plot the number of pixels with respect to their standard deviation, obtained by repeating the reconstruction for $20$ times. The vertical axis is in $\log_{10}$ scale. The inset displays the standard deviation at each pixel location. (a) Downsampling in the standard computational basis, i.e. without using the Hadamard gates. (b) Downsampling in the $X$ basis, i.e. with the Hadamard gates, as reported in \cref{fig:Algorithm1}. Statistical fluctuations reduce when using the Hadamard, while keeping the same number of shots.}
\end{figure*}

In \cref{fig:ComparisonHadamardExample}, we simulate the downsampling of a one-dimensional array pattern, with and without using the Hadamard gates. Without the Hadamard, \cref{alg:Algorithm1} reproduces the original input, but with lowered contrast. Moreover, non-symmetrical boundary artifacts occur for a ladder, i.e. linear and monotonic, pattern. This effect increases with the number of discarded qubits, i.e. inversely with the size of $\IQFT_1$. In both cases, the Hadamard improves the quality of the output: it regularizes the contrast, while removing the artifacts in the ladder pattern. In \cref{fig:HadamardComparison} we show that, for a generic $512\times512$ image, the Hadamard also reduces the statistical fluctuations in the output reconstruction, while keeping the same number of shots.

\section{\label{app:B}JAYNES-CUMMINGS HARDWARE ENCODING}
In this section we show how to obtain the state of \cref{eq:OneHotEncoding} from the multimode and multiatom Jaynes-Cummings model introduced in \citep{art:Emary_MultiJC, art:Wickenbrock_MultiJC}, also known as multimode Tavis-Cummings model \citep{art:Tavis}. We adopt units with $\hbar = 1$. We work in the interaction picture and in the rotating wave approximation \citep{book:Gerry}, in which the free Hamiltonian reads
\begin{equation}
	H_F = \frac{1}{2} \sum_{m,n=0}^{N-1} \omega_0 \sigma^{(3)}_{mn} + \int dk \omega(k) a^+(k) a(k) \ , 
\end{equation}	
with $\sigma^{(3)}_{mn} = \ket{1}_{mn}\bra{1}_{mn} - \ket{0}_{mn}\bra{0}_{mn}$, while the interaction term yields
\begin{equation}
	H_{I} = \int dk \sum_{m,n=0}^{N-1} \gamma^*_{mn}(k) a(k)\sigma_{mn}^{+} + \text{ H.c.} 
\end{equation}
The total Hamiltonian of the system reads
\begin{equation}
	H_{JC} = H_F + \lambda H_{I} \ ,
\end{equation}
where $\lambda$ is an overall coupling constant.

Prepare the system in the state $\dket{I} = \ket{\alpha_\psi} \otimes \ket{b}$. Since $[H_F, H_{I}]=0$, the evolution of $\dket{I}$ is completely driven by the unitary operator generated by $H_{I}$, so that
\begin{equation}
	U_t = \Unit + i t \lambda H_{I} + \mathcal{O}(\lambda^2) \ .
\end{equation}
The final state $\dket{F} = U_{t}\ket{I}$ then reads
\begin{equation}
	\dket{F} = \ket{\alpha_\psi} \otimes \ket{b} + i t \lambda H_{I} \ket{I} + \mathcal{O}(\lambda^2) \ . 
\end{equation}
We neglect the first term by post-selectively discarding any completely black image at the output. Then, the leading order contribution becomes dominant and the evolution is solely controlled by the Jaynes-Cummings Hamiltonian
\begin{equation}
	\dket{F} \simeq  it \lambda \int dk \sum_{m,n=0}^{N-1} \gamma^*_{mn}(k) a(k)\sigma_{mn}^{+} \ket{\alpha_\psi} \otimes \ket{b} \ ,
\end{equation}
where the Hermitian conjugate is annihilated on $\ket{b}$. For the multimode coherent state 
\begin{equation}
	D(\alpha_\psi)\ket{\varnothing} \simeq \prod_{k} \exp( \widetilde{\psi}(k) a^\dagger(k) + \text{ H.c.}) \ ,
\end{equation}
then $a(k)\ket{\alpha_\psi} = \alpha \widetilde{\psi}(k) \ket{\alpha_\psi}$. A projection on the coherent state gives $\ket{\Omega} = \langle \alpha_\psi\dket{F}$. By substituting the Fourier transform of $\psi$ and $g^*_{mn}$, we get
\begin{equation}
	\ket{\Omega} = it \lambda \alpha \sum_{m,n=0}^{N-1} \int dr \psi(r)  g^*_{mn}(r) \sigma^+_{mn} \ket{b} \ ,
\end{equation}
which is precisely the result shown in  \cref{eq:OneHotEncoding,eq:EncodingCoefficients}.

\FloatBarrier
\bibliography{refs.bib}
\end{document}